\documentclass[aps,prl,twocolumn,epsfig,groupedaddress]{revtex4}

\usepackage{epsfig}
\usepackage{graphics}
\begin{document}

\def\lsim{\raise0.3ex\hbox{\,$<$\kern-0.75em\raise-1.1ex\hbox{$\sim$}\,}}
\def\gsim{\raise0.3ex\hbox{\,$>$\kern-0.75em\raise-1.1ex\hbox{$\sim$}\,}}

\title{Measuring The Collective Flow With Jets}

\author{N\'estor Armesto, Carlos A. Salgado and Urs Achim Wiedemann}
\address{Department of Physics, CERN, Theory Division, CH-1211 
Gen\`eve 23, Switzerland}

\date{\today}

\begin{abstract}
In nucleus--nucleus collisions, high-$p_T$ partons interact with
a dense medium, which possesses strong collective flow components.
Here, we demonstrate that the resulting medium-induced gluon 
radiation does not depend solely on the energy density of the 
medium, but also on the collective flow. Both components cannot be 
disentangled on the basis of leading hadron spectra, but the 
measurement of particle production associated to high-$p_T$ 
trigger particles, jet-like correlations and jets, allows for 
their independent characterization. In particular, we show that 
flow effects lead to a characteristic breaking of the rotational 
symmetry of the average jet energy and jet multiplicity distribution 
in the $\eta \times \phi$-plane. We argue that data on the medium-induced 
broadening of jet-like particle correlations in Au+Au collisions
at RHIC provide a first evidence for a significant distortion of 
parton fragmentation due to the longitudinal collective flow.
\end{abstract}
\maketitle
 \vskip 0.3cm


One of the most striking generic features of nucleus--nucleus
collisions at the Relativistic Heavy Ion Collider RHIC is
the extent to which they support a hydrodynamic 
interpretation~\cite{Ackermann:2000tr,Adcox:2002ms,Adler:2003kt,Kolb:2003dz}.
For transverse momentum up to $p_T \lsim 2$ GeV, the hadronic
transverse momentum spectra, their azimuthal asymmetry and 
particle species dependence, as well as the shape of rapidity 
distributions can be accounted for by modelling the medium as an 
ideal fluid, supplemented by a thermal freeze-out 
condition~\cite{Kolb:2003dz,Kolb:2001qz,Teaney:2000cw,Hirano:2003yp,Retiere:2003kf}.
Although we still lack a microscopic understanding of why
hydrodynamics is successful~\cite{Molnar:2001ux}, this success is 
regarded as strong evidence~\cite{Gyulassy:2004zy}
that the medium produced in nucleus--nucleus collisions 
at RHIC equilibrates efficiently and builds up significant 
position--momentum correlations, i.e. a flow field $u_\mu(x)$. 
That bulk properties of nucleus-nucleus collisions are
calculable from the energy--momentum tensor 
\begin{equation}
  T^{\mu\nu}(x) = \left(\epsilon + p \right)\, u^{\mu}\, u^{\nu}\, 
                  - p\, g^{\mu \nu}
  \label{eq1} 
\end{equation}
of an ideal fluid also further supports the expectation that
nucleus--nucleus collisions give access to the equation of state 
$p = p(\epsilon,T,\mu_B)$ of dense QCD matter.

At collider energies, the production of high-$p_T$ hadrons and
jets provides a novel independent characterization of the 
medium produced in nucleus--nucleus collisions. This is so since 
the gluon radiation off parent partons is sensitive to the 
interaction between the partonic projectile and the 
medium~\cite{Gyulassy:1993hr,Baier:1996sk,Zakharov:1997uu,Wiedemann:2000za,Gyulassy:2000er,Wang:2001if}. In particular, quenched high-$p_T$ hadroproduction
is sensitive to the transport coefficient $\hat{q}$, which is 
proportional to the density of scattering centres and characterizes 
the squared average momentum transfer from the medium to the hard
parton per unit path length. This transport coefficient is related 
to the energy density of the medium~\cite{Baier:2002tc},
\begin{equation}
  \hat{q} \left[{\rm GeV}^2/{\rm fm} \right]\, 
  = \, c \, \epsilon^{3/4} \left[({\rm GeV}/{\rm fm}^3)^{3/4} \right]\, .
  \label{eq2} 
\end{equation}
Here, $c$ is a proportionality constant of order unity whose value 
can be established in model calculations~\cite{Baier:2002tc} or in 
comparison with experimental data~\cite{Helitoappear}. 

\begin{figure}[h]

\begin{center}
\includegraphics[width=5.3cm,angle=-90]{carlosflows.epsi}
\end{center}

\begin{center}
\includegraphics[width=8.5cm]{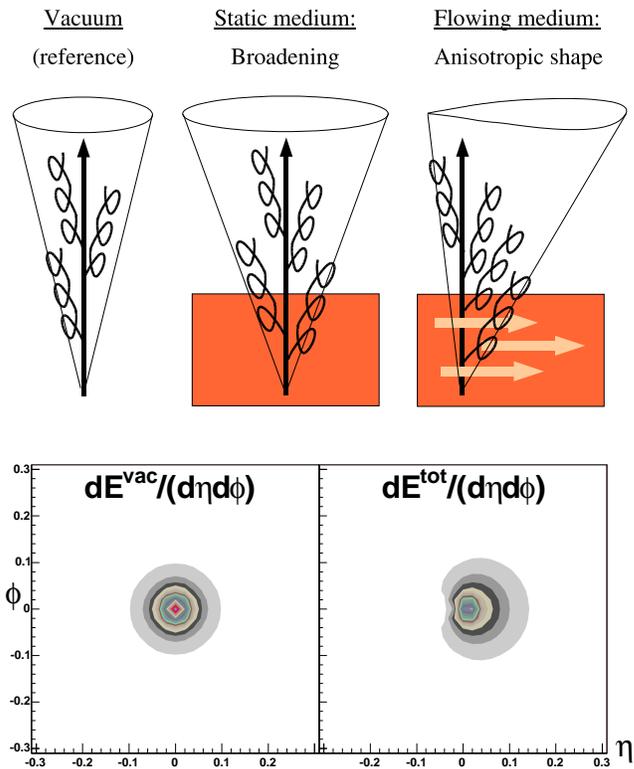}
\end{center}

\vspace{-0.5cm}
\caption{Upper part: sketch of the distortion
of the jet energy distribution in the presence of a medium
with or without collective flow. Lower part: calculated
distortion of the jet energy distribution (\protect\ref{eq5})
in the $\eta \times \phi$-plane for a 100 GeV jet. The right
hand-side is for an 
average medium-induced radiated energy of 23 GeV and equal
contributions from density and flow effects, $\mu = q_0$. 
Scales of the contour plot are visible from Fig.~\ref{fig2}.
}\label{fig1}
\end{figure}

The main purpose of this letter is to demonstrate that the sensitivity 
of parton energy loss is not limited to the energy density $\epsilon$ 
of the produced matter, but extends to the strength and direction 
of the collective flow field $u_\mu(x)$ in (\ref{eq1}). Recent works 
on parton energy loss account for the rapid decrease of the density 
of scattering centres caused by collective 
expansion~\cite{Baier:1998yf,Gyulassy:2000gk,Salgado:2002cd}.  
This effect can be absorbed in a redefinition of the static 
transport coefficient~\cite{Salgado:2002cd}. Here we go beyond
these formulations by taking into account that, in heavy-ion
collisions, hard partons are not produced in a rest frame in 
which the momentum transfer from the medium is isotropic in the
plane transverse to the direction of parton propagation 
(``isotropic rest frame''), see
Fig.~\ref{fig1}. Rather, they interact with a medium which
generically shows collective flow components in this transverse plane. 
To illustrate the consequences, we consider the Gyulassy--Wang 
model~\cite{Gyulassy:1993hr}, 
which idealizes the medium as a collection of coloured Yukawa-type 
scattering potentials $a({\bf q})$ with Debye screening mass $\mu$. 
For a coloured test particle in an isotropic rest frame, the 
average momentum transfer per scattering centre is $\mu$.
In the presence of collective flow, however, the hard partonic
projectile interacts with Lorentz-boosted scattering centres
which can be modelled by a momentum shift ${\bf q}_0$.
This shift is proportional to the flow field component which 
points transverse to the direction of parton propagation, 
\begin{equation}
  \vert a({\bf q}) \vert^2
  = \frac{\mu^2}{\pi \left[ ({\bf q} - {\bf q}_0)^2 + \mu^2 \right]^2}\, .
  \label{eq3}
\end{equation}

We have calculated the medium-induced radiation of gluons with energy 
$\omega$ and transverse momentum ${\bf k}$, emitted from a highly 
energetic parton that propagates over a finite path length $L$ in a 
medium of density $n_0$ with collective 
motion. To first order in opacity, we 
find~\cite{Wiedemann:2000za,Gyulassy:2000er,Salgado:2003gb} 
\begin{eqnarray}
 \omega \frac{dI^{\rm med}}{d\omega\, d{\bf k}} &=& \frac{\alpha_s}{(2\pi)^2}
 \frac{4\, C_R\, n_0}{\omega} \, 
 \int d{\bf q}\, \vert a({\bf q})\vert^2\, 
 \frac{ {\bf k}\cdot {\bf q}}{{\bf k}^2}
 \nonumber \\
 && \times
 \frac{-L \frac{({\bf k} + {\bf q})^2}{2\omega} + 
       \sin\left(   L \frac{({\bf k} + {\bf q})^2}{2\omega}\right)}
     {  \left[({\bf k} + {\bf q})^2 / 2\omega\right]^2} \, .
 \label{eq4}
\end{eqnarray}
In the absence of a medium, the parton fragments according to the vacuum 
distribution $I^{\rm tot} = I^{\rm vac}$. The radiation spectrum (\ref{eq4}) 
characterizes the medium modification of this distribution
$  \omega\frac{dI^{\rm tot}}{d\omega\, d{\bf k}}
  = \omega\frac{dI^{\rm vac}}{d\omega\, d{\bf k}}
  + \omega\frac{dI^{\rm med}}{d\omega\, d{\bf k}}$ .
From this, we calculate distortions of jet energy and 
jet multiplicity distributions~\cite{Salgado:2003rv}.
Information about $I^{\rm vac}$ is obtained
from the energy fraction of the jet contained in a subcone of radius 
$R = \sqrt{\eta^2 + \phi^2}$,
\begin{eqnarray}
  &&\rho_{\rm vac}(R) \equiv \frac{1}{N_{\rm jets}} \sum_{\rm jets}
  \frac{E_T(R)}{E_T(R=1)}
  \nonumber\\
  && \quad = 1 -  \frac{1}{E_T} \int d\omega \int^\omega d{\bf k}\, 
  \Theta\left(\frac{k}{\omega} - R\right)\, 
  \omega\frac{dI^{\rm vac}}{d\omega\, d{\bf k}}\, .
  \label{eq5}
\end{eqnarray}
For this jet shape, we use the parametrization~\cite{Abbott:1997fc} 
of the Fermilab $D0$ Collaboration for jet energies in the range 
$\approx 50 < E_t < 150$ GeV and opening cones $0.1 < R < 1.0$. 
We remove the unphysical singularity of this parametrization 
for $R \to 0$ by smoothly interpolating with a polynomial ansatz
for $R < 0.04$ to $\rho(R=0) = 0$. We then calculate from 
eq.~(\ref{eq4}) the modification~\cite{Salgado:2003rv} of 
$\rho_{\rm vac}(R)$ caused by the energy density and collective flow 
of the medium. To do so, we transform the gluon emission
angle ${\rm arcsin}\left(k/\omega\right)$ in (\ref{eq4}) to jet 
coordinates $\eta$, $\phi$,
\begin{equation}
  k\, dk\, d\alpha = \omega^2\, \frac{\cos\phi}{\cosh^3\eta}\, 
  d\eta\, d\phi\, ,
  \label{eq6}
\end{equation}
where $\alpha$ denotes the angle between the transverse gluon 
momentum ${\bf k}$ and the collective flow component ${\bf q}_0$. 
In what follows, we mainly focus on changes of the jet shape due to 
longitudinal collective flow effects where the directed 
momentum transfer ${\bf q}_0$ points along the beam direction.
The sensitivity of jets and leading hadron spectra to other 
collective flow components will be discussed elsewhere~\cite{Nestornew}.

To specify input values for the momentum transfer from the medium,
we make the following considerations. First, for a given density 
$n_0$ of scattering centres, the transport coefficient 
is given as $\hat q \simeq n_0\, \mu^2$, see Ref.~\cite{Salgado:2003gb}. 
Thus, according to (\ref{eq2}), the hard parton suffers a momentum 
transfer that is monotonously increasing with the pressure in the medium, 
$n_0\, \mu^2 \propto p^{3/4}$ and which tests the components $T^{\perp\perp}$
and $T^{zz}$ ($z$ parallel to the beam) of the energy momentum tensor 
(\ref{eq1}). In the presence of a longitudinal Bjorken-type flow field
$u^{\mu} = \left(1,\vec{\beta}\right)/\sqrt{1 - \beta^2}$, the 
longitudinal flow component increases from $T^{zz} = p$ to
$T^{zz} = p + \Delta p$, where
$\Delta p = (\epsilon + p) u^z\, u^z = 4\, p\, \beta^2/(1-\beta^2)$ 
for the equation of state of an ideal gas, $\epsilon = 3\, p$. 
For a rapidity difference $\eta = 0.5, 1.0, 1.5$ between the rest 
frame, which is longitudinally comoving with the jet, and the rest
frame of the medium, this corresponds to an increase of the 
component $T^{zz}$ by a factor $1, 5, 18$, respectively. 
We expect that the collective flow component $q_0$ rises monotonously
with the flow-induced $\Delta p$, as $\mu$ does with $p$. 
This suggests that $q_0$ lies in the parameter 
range $q_0 \gsim \mu$. 

In Fig.~\ref{fig1}, we show the medium-modified jet shape for 
a jet of total energy $E_T = 100$ GeV. To test the sensitivity
of this energy distribution to collective flow, we have chosen
a rather small directed flow component, $q_0 = \mu$. The
effective coupling constant in (\ref{eq3}), $n_0\, L\, \alpha_s\, 
C_R = 1$, the momentum transfer per scattering centre $\mu = 1$ GeV,
and the length of the medium $L = 6$ fm were adjusted such that
an average energy $\Delta E_T = \int d\omega 
\frac{dI^{\rm med}}{d\omega} = 23$ GeV is redistributed by
medium-induced gluon radiation. Previous studies indicate that
this value of $\Delta E_T$ is a conservative estimate for the
modification of jets produced in Pb+Pb collisions at the 
Large Hadron Collider LHC~\cite{Salgado:2003rv}. 
Despite these conservative estimates, the contour plot of the
jet energy distribution in  Fig.~\ref{fig1} displays marked 
medium-induced deviations. First, the jet structure broadens
because of the medium-induced Brownian motion of the partonic jet 
fragments in a dense medium~\cite{Salgado:2003gb}. Second, the
jet shape shows a marked rotational asymmetry in the 
$\eta \times \phi$-plane, which is characteristic of the presence 
of a collective flow field.

We note that for each single jet, the 
$\eta\times\phi$-rotation symmetry is broken even in the absence of a medium. 
First, any finite multiplicity distribution of a rotationally 
symmetric sample breaks the symmetry by terms proportional to 
$1/\sqrt{N}$. In principle, this can be corrected for by
the method used to analyze elliptic flow~\cite{Borghini:2000sa}.
Second, the $k_T$-ordering of the final
state DGLAP parton shower implies that the first parton splitting 
in the shower contains significantly more transverse momentum 
than the following ones, thus leading to a dynamical 
asymmetry in the $\eta \times \phi$-plane. Both these
effects lead to a symmetry breaking in a {\it random} direction in 
the $\eta \times \phi$-plane; thus rotational symmetry is restored
in sufficiently large jet samples. A third source of asymmetry in the 
$\eta \times \phi$ plane is not random but related to the Jacobian 
in (\ref{eq6}). We have checked that the resulting asymmetry is
$< 10 \%$ for $R < 0.3$ but can become sizeable for larger
jet cones. Most importantly, distributions that are
rotationally symmetric in $\alpha$ are elongated by the Jacobian 
(\ref{eq6}) in the $\phi$-direction. This choice of coordinates
reduces the effect of $\eta$-broadening due to longitudinal flow,
but can be corrected for analytically. 
%
\begin{figure}[h]\epsfxsize=9.7cm
\centerline{\epsfbox{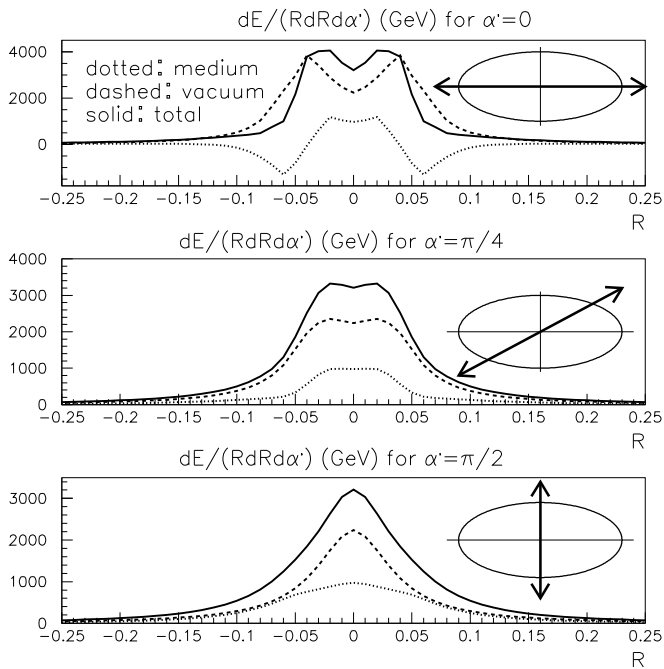}}
\vspace{-0.5cm}
\caption{Jet energy distribution for a sample of jets for which
the medium was moving with equal probability in the positive and
negative beam direction. 
}\label{fig2}
\end{figure}
%

For collisions of identical nuclei, jet samples centered around mid 
(momentum) rapidity have to be symmetric with respect to
$\eta \to -\eta$. Thus, while each parent parton may experience on 
average a significant collective flow, the direction of the
oriented momentum transfer points with equal probability in
the positive or negative beam direction, $+ {\bf q}_0$ or
$- {\bf q}_0$, respectively. In Fig.~\ref{fig2}, we show the
average jet energy distribution for the resulting symmetrized 
jet sample. Jet samples are symmetrized by identifying 
the calorimetric centres of every jet in the sample, thus
mimicking the experimental procedure. Since in
the presence of collective flow the energy distribution of each 
jet is asymmetric with respect to its calorimetric centre, 
this can result in an energy distribution with a double-hump shape,
as seen in Fig.~\ref{fig2}. However, details of the shape of the
energy distribution may be subject to significant uncertainties in 
the parametrization of (\ref{eq5}) and the calculation of 
(\ref{eq4}) at small angles. In contrast, the asymmetric broadening of 
the energy distribution in the $\eta \times \phi$-plane is a
generic characteristics of collective flow.

%
\begin{figure}[t!]\epsfxsize=8.7cm
\centerline{\epsfbox{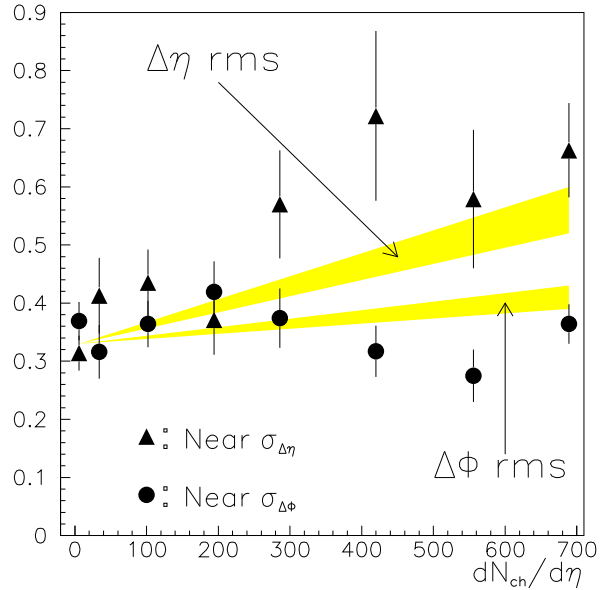}}
\caption{The width in azimuth and rapidity of the near-side 
distribution of charged hadrons associated to high-$p_T$
trigger particles of transverse momentum 
$4\, {\rm GeV} <  p_T < 6\, {\rm GeV}$ in Au+Au collisions
at $\sqrt{s_{NN}} = 200$ GeV. Black points are preliminary data 
from the STAR collaboration~\cite{FuqiangWang}. 
The band represents our
calculation for longitudinal flow fields in the range 
$2< q_0/\mu < 4$, see text for further details.   
}\label{fig3}
\end{figure}
%
While the calculation of medium-induced gluon radiation is 
most reliable for calorimetric measurements, it also provides a
framework for the discussion of medium-modified multiplicity 
distributions. In particular, we have checked that the azimuthal 
asymmetries seen in Figs.~\ref{fig1} and ~\ref{fig2} also 
persist on the level of leading near-side two-particle correlations.
This allows us to test our formalism versus preliminary data 
of the STAR Collaboration~\cite{FuqiangWang,Wang:2004kf}, 
which measured the widths of the $\eta$- and 
$\phi$-distributions of produced hadrons associated to trigger particles
of transverse momentum $4\, {\rm GeV} <  p_T < 6\, {\rm GeV}$.
As a function of centrality of the collision, the $\phi$-distribution
does not change within errors, while the $\eta$-distribution shows
a significant broadening, see Fig.~\ref{fig3}. In our calculation 
of this effect, we
have used the width of the jet-like correlation in p+p collisions
to characterize the vacuum contribution. The energy of the parent
parton was fixed to 10 GeV. We 
have chosen a rather small in-medium path length of $L = 2$ fm
to account for the fact that high-$p_T$ trigger particles tend
to correspond to parent partons produced near the surface. 
We then calculated the asymmetry of the broadening in 
$\Delta \eta$ and $\Delta \phi$ by varying the average momentum
transfer between $\mu  = 0.7$ and $\mu = 1.4$ GeV, and the
size of the collective flow component between $q_0/\mu = 2$
and $q_0/\mu = 4$. The results thus obtained for central Au+Au
collisions were extrapolated to peripheral ones by a straight line
and are represented by the band in Fig.~\ref{fig3}.  
Numerical uncertainties in applying calculations of parton energy
loss to transverse hadron momenta $p_T < 10$ GeV are significant
and have been discussed repeatedly~\cite{Salgado:2003gb}.  
However, the origin of the 
angular broadening of jet-like particle correlations is essentially 
kinematic, being determined by the ratio between the momentum
transfer from the medium and the energy of the escaping particle;
hence, the result in Fig.~\ref{fig3} should not depend strongly 
on the details of our calculation. 

Based on the observation that the ratio $q_0/\mu = 4$ can account 
for the tendency in the preliminary STAR data of Fig.~\ref{fig3},
we conjecture a picture of the space-time distribution of hard 
processes in nucleus--nucleus collisions. The ratio $q_0/\mu = 4$
corresponds to a boost of the energy--momentum tensor (\ref{eq1}) 
by approximately one unit in rapidity $\Delta \eta$. 
This indicates that in Au+Au collisions at RHIC the hard parent 
partons of $4\, {\rm GeV} <  p_T < 6\, {\rm GeV}$ trigger particles 
are produced on average distance $\Delta \eta$ away from the 
part of the medium that is locally 
longitudinally comoving with their rest frame.

In summary, we have established that jet energy distributions and
jet-like particle correlations are sensitive to the density of
the medium and its position--momentum correlations such as 
a collective flow field. Remarkably, comparing
a four-fold larger flow component $q_0/\mu = 4$ with $q_0/\mu = 0$,
we find that the average parton energy loss more than doubles. 
This indicates that the energy density produced in the medium can
be overestimated significantly if flow effects are ignored. In
our view, this is not the only reason why the flow effect discussed 
here will play in important role in further comparisons of parton 
energy loss with data on jet quenching. Other motivations include 
the novel possibility to determine the space-time distribution of 
hard processes in the medium e.g. by refined studies of the interplay 
of parton energy loss and hydrodynamic simulations~\cite{Hirano:2002sc}.
Moreover, the formulation of parton energy loss given here implies
that partons lose less energy if they escape along trajectories 
parallel to the transverse flow field generated in nucleus-nucleus
collisions. Compared to calculations for a static medium, 
this enhances the parton energy loss contribution to elliptic 
flow~\cite{Nestornew}. 

We acknowledge helpful discussions with Rolf Baier, J\"urgen Schukraft 
and Fuqiang Wang. We thank Fuqiang Wang for providing us with the
data of Ref.~\cite{FuqiangWang}.

%


\begin{thebibliography}{9}
%
\bibitem{Ackermann:2000tr}
K.~H.~Ackermann {\it et al.}  [STAR Collaboration],
Phys.\ Rev.\ Lett.\  {\bf 86} (2001) 402.
%
\bibitem{Adcox:2002ms}
K.~Adcox {\it et al.}  [PHENIX Collaboration],
Phys.\ Rev.\ Lett.\  {\bf 89} (2002) 212301.
%
\bibitem{Adler:2003kt}
S.~S.~Adler {\it et al.}  [PHENIX Collaboration],
Phys.\ Rev.\ Lett.\  {\bf 91} (2003) 182301.
%
\bibitem{Kolb:2003dz}
P.~F.~Kolb and U.~Heinz,
arXiv:nucl-th/0305084.
%
\bibitem{Kolb:2001qz}
P.~F.~Kolb, U.~W.~Heinz, P.~Huovinen, K.~J.~Eskola and K.~Tuominen,
Nucl.\ Phys.\ A {\bf 696} (2001) 197
[arXiv:hep-ph/0103234].
%
\bibitem{Teaney:2000cw}
D.~Teaney, J.~Lauret and E.~V.~Shuryak,
Phys.\ Rev.\ Lett.\  {\bf 86} (2001) 4783.
%
\bibitem{Hirano:2003yp}
T.~Hirano and Y.~Nara,
Phys.\ Rev.\ C {\bf 68} (2003) 064902.
%
\bibitem{Retiere:2003kf}
F.~Retiere and M.~A.~Lisa,
arXiv:nucl-th/0312024.
%
\bibitem{Molnar:2001ux}
D.~Molnar and M.~Gyulassy,
Nucl.\ Phys.\ A {\bf 697} (2002) 495.
%
\bibitem{Gyulassy:2004zy}
M.~Gyulassy and L.~McLerran,
arXiv:nucl-th/0405013.
%
\bibitem{Gyulassy:1993hr}
M.~Gyulassy and X.~N.~Wang,
Nucl.\ Phys.\ B {\bf 420} (1994) 583.
%
\bibitem{Baier:1996sk}
R.~Baier, Y.~L.~Dokshitzer, A.~H.~Mueller, S.~Peigne and D.~Schiff,
Nucl.\ Phys.\ B {\bf 484} (1997) 265.
%
\bibitem{Zakharov:1997uu}
B.~G.~Zakharov,
JETP Lett.\  {\bf 65} (1997) 615.
%
\bibitem{Wiedemann:2000za}
U.~A.~Wiedemann,
Nucl.\ Phys.\ B {\bf 588} (2000) 303.
%
\bibitem{Gyulassy:2000er}
M.~Gyulassy, P.~Levai and I.~Vitev,
Nucl.\ Phys.\ B {\bf 594} (2001) 371.
%
\bibitem{Wang:2001if}
X.~N.~Wang and X.~f.~Guo,
Nucl.\ Phys.\ A {\bf 696} (2001) 788.
%
\bibitem{Baier:2002tc}
R.~Baier,
Nucl.\ Phys.\ A {\bf 715} (2003) 209.
%
\bibitem{Helitoappear}
K.~Eskola, H.~Honkanen, C.~A.~Salgado, and U.~A.~Wiedemann,
in preparation.
%
\bibitem{Baier:1998yf}
R.~Baier, Y.~L.~Dokshitzer, A.~H.~Mueller and D.~Schiff,
Phys.\ Rev.\ C {\bf 58} (1998) 1706.
%
\bibitem{Gyulassy:2000gk}
M.~Gyulassy, I.~Vitev and X.~N.~Wang,
Phys.\ Rev.\ Lett.\  {\bf 86} (2001) 2537.
%
\bibitem{Salgado:2002cd}
C.~A.~Salgado and U.~A.~Wiedemann,
Phys.\ Rev.\ Lett.\  {\bf 89} (2002) 092303
%
\bibitem{Salgado:2003gb}
C.~A.~Salgado and U.~A.~Wiedemann,
Phys.\ Rev.\ D {\bf 68} (2003) 014008.
%
\bibitem{Salgado:2003rv}
C.~A.~Salgado and U.~A.~Wiedemann,
arXiv:hep-ph/0310079, Phys. Rev. Lett. in press.
%
\bibitem{Abbott:1997fc}
B.~Abbott, M.~Bhattacharjee, D.~Elvira, F.~Nang and H.~Weerts  [D0 Coll.],
FERMILAB-PUB-97-242-E
%
\bibitem{Nestornew}
N.~Armesto, C.~A.~Salgado and U.~A.~Wiedemann, in preparation.
%
\bibitem{Borghini:2000sa}
N.~Borghini, P.~M.~Dinh and J.~Y.~Ollitrault,
Phys.\ Rev.\ C {\bf 63} (2001) 054906.
%
\bibitem{FuqiangWang}
F.~Wang for the STAR Collaboration, talk at QM'04 Conference,
Oakland, 11-17 Jan 2004, http://www.lbl.gov/nsd/qm2004/.
%
\bibitem{Wang:2004kf}
F.~Wang  [STAR Collaboration],
arXiv:nucl-ex/0404010.
%
\bibitem{Hirano:2002sc}
T.~Hirano and Y.~Nara,
Phys.\ Rev.\ C {\bf 66} (2002) 041901.
%
\end{thebibliography}
\end{document}